\documentclass{article}
\usepackage{LaThuileFPSpro,epsfig}
\begin{document}
\title{ 
  ELECTROWEAK MEASUREMENTS AT THE TEVATRON
  }
\author{
  Kristian Harder\\
  {\em Rutherford Appleton Laboratory} \\
  for the CDF and D\O~Collaborations\\
  }
\maketitle

\baselineskip=11.6pt

\begin{abstract}
  The increasing size of the data samples recorded by the CDF and
  D\O~experiments at the Tevatron enables studies of a wide range of
  processes involving the electroweak bosons W and Z. Single boson
  production is now looked at in terms of differential cross sections
  such as rapidity or transverse momentum dependence. Diboson production
  cross-sections are several orders of magnitude smaller than single boson
  production cross-sections, but all combinations W$\gamma$, Z$\gamma$,
  WW and WZ have been observed. ZZ production is expected at a rate just
  below the observation threshold with current data sample sizes, but this
  channel is expected to be accessible to the Tevatron experiments soon.
\end{abstract}
\newpage
\section{Introduction}
Leptonic final states of W and Z boson decays exhibit a very clear
experimental signature and pave the way for precision tests of the
Standard Model beyond leading order and possible detection on non
Standard Model contributions. These measurements can provide strong
constraints to parton density functions.

The hadronic collision data recorded by the Tevatron experiments CDF
and D\O~as of early 2007 amount to more than 2$\,$fb$^{-1}$ per
experiment, about 1$\,$fb$^{-1}$ each of which have been made
available for electroweak physics analysis so far. While measurements
of the total production cross-section of single W or Z bosons were
already performed on much smaller size samples, the current data set
allows for a much more in-depth analysis of the production process by
measuring differential cross-sections. Also, most diboson production
processes are now experimentally accessible despite their lower
cross-section. In the following, we will summarise the typical W and Z
selection procedure applied by the CDF and D\O~experiments, and then
present recent electroweak results made available by both
collaborations.

\section{W and Z reconstruction}
Both CDF and D\O~follow a fairly standard path for boson
reconstruction, with only minor variations e.g.~in cut thresholds
between the different experiments or different analyses from the same
collaboration.

Electrons are identified from calorimeter clusters that pass shower
shape requirements and have a transverse momentum in excess of
typically 20$\,$GeV. Isolation cuts are applied to remove background
from fake electrons and electrons in jets.  Both D\O~and CDF perform
their reconstruction separately in the central barrel calorimeters and
their forward calorimeters, while not using data from the intermediate
region where modelling of the detector response is more difficult.

Muon reconstruction is based on signals identified in the muon
detectors or calorimeters. In cases where efficiency is most
important, CDF also includes tracks without associated muon or
calorimeter signal in their muon selection. A transverse momentum
threshold around 20 GeV is applied, and the muon candidates are
required to be isolated in the tracking system and/or calorimeter to
remove background from muons from heavy quark decay. The
pseudorapidity coverage of muons used in the CDF analyses is
restricted to $|\eta|\leq$1.1--1.2, whereas D\O~has muons in the range
up to $|\eta|\leq$2 at their disposal.

Tau leptons are not treated separately. Leptonically decaying taus are
implicitly included in the electron and muon selections.

Leptonic W boson decays involve a neutrino, which is exploited for the
reconstruction by requiring missing transverse energy of typically at
least 20$\,$GeV in candidate events. CDF requires the missing momentum
vector to be isolated.

\section{Differential Z cross-sections}

Leptonic Z decays can be reconstructed fully and therefore provide the
laboratory of choice for studying the intricacies of single
electroweak boson production processes, despite the cross-section
being an order of magnitude smaller than that of W production.  The
sample of reconstructed Z bosons collected at the Tevatron is large
enough to investigate the dependence of the production cross-section
on quantities such as Z rapidity and Z transverse momentum
distribution.

\begin{figure}
\begin{center}
\epsfig{file=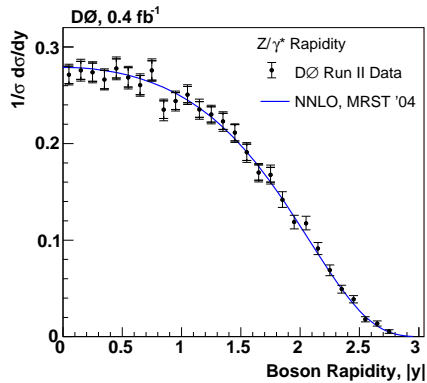, height=5cm}
\caption{\label{fig-d0zrap}\it Z boson rapidity distribution observed by D\O.
  This recently published 0.4$\,$fb$^{-1}$ result\cite{cite-d0zrap} is expected to be followed
  up by a 1$\,$fb$^{-1}$ result soon.}
\end{center}
\end{figure}

The Z rapidity distribution is especially interesting in the forward
region, where it provides constraints for parton density functions at
low momentum fraction $x$ and large momentum transfer $Q^2$, as well
as at large $x$. Both D\O\cite{cite-d0zrap} and CDF\cite{cite-cdfzrap}
do this measurement in the Z$\to$ee channel due to the larger $\eta$
coverage of the calorimeter compared to the muon system, $|\eta|<$3.2
at D\O, $|\eta|<$2.8 at CDF.  The observed distributions are compared
to NNLO predictions (MRST '04, CTEQ6.1) and found to be in good
agreement, as demonstrated for example in Fig.~\ref{fig-d0zrap}.

D\O~also measures the transverse momentum distribution of Z
bosons\cite{cite-d0zpt}. This distribution is very sensitive to higher
order effect because there is no leading order contribution to Z
transverse momentum. Prediction of this distribution requires
resummation. Although the current revision of the measurement is not
yet able to distinguish between different calculations, good agreement
is found with the available predictions, and the sensitivity of the
analysis to model differences is expected to be improved on a short
time scale.

\section{Diboson production}

\begin{table}
\resizebox{\textwidth}{!}{
\begin{tabular}{|c|c|r|r|r|r|}
\hline
 & exp. & \multicolumn{1}{c|}{sample} & evts & \multicolumn{1}{c|}{prediction} & \multicolumn{1}{c|}{measured cross-section} \\
        &      &                             &      & \multicolumn{1}{c|}{(SM, in pb)} & \multicolumn{1}{c|}{(pb)} \\
\hline
                                      & CDF & 1.1$\,$fb$^{-1}$ & 855 & $19.3\pm1.4$ & 19.11$\pm$1.04(stat)$\pm$2.40(syst)$\pm$1.11(lum)\\
\raisebox{1.5ex}[0ex][0ex]{W$\gamma$ ($\mu$)} & D\O & 1$\,$fb$^{-1}$   & 245 & $3.21\pm0.08$ & 3.21$\pm$0.49(stat+syst)$\pm$0.20(lum)\\
W$\gamma$ ($e$)                       & D\O & 1$\,$fb$^{-1}$ & 389 & $3.21\pm0.08$& 3.12$\pm$0.49(stat+syst)$\pm$0.19(lum)\\
                                      & CDF & 1.1$\,$fb$^{-1}$ & 390 & $4.7\pm0.4$ & 4.9$\pm$0.3(stat)$\pm$0.3(syst)$\pm$0.3(lum)\\
\raisebox{1.5ex}[0ex][0ex]{Z$\gamma$} & D\O & 1$\,$fb$^{-1}$   & 387 & $4.2\pm0.4$ & 4.51$\pm$0.37(stat+syst)$\pm$0.27(lum)\\
WW                                    & CDF & 0.8$\,$fb$^{-1}$ & 95 & $12.4\pm0.8$ & 13.6$\pm$2.3(stat)$\pm$1.6(syst)$\pm1.2$(lum)\\
                                      & CDF & 1.1$\,$fb$^{-1}$ & 16 & $3.7\pm0.3$ & 5.0$^{+1.8}_{-1.4}$(stat)$\pm$0.4(syst)\\
\raisebox{1.5ex}[0ex][0ex]{WZ}        & D\O & 1$\,$fb$^{-1}$   & 12 & $3.7\pm0.3$ & 4.0$^{+1.9}_{-1.5}$(stat+syst)\\
ZZ                                    & CDF & 1.1$\,$fb$^{-1}$ & 1 & $1.4\pm0.1$ & $<$3.8 (95\% C.L.)\\
\hline
\end{tabular}
}
\caption{\label{tab-xsec}\it Overview of diboson production cross-section measurements discussed in this document.  Predictions are as quoted in the respective analysis write-up. Analysis of different kinematic regions leads to different cross-section predictions for the same channel. In particular, D\O~uses a very stringent FSR veto cut in their W$\gamma$ analysis, whereas CDF does not.}
\end{table}

Production processes of gauge boson pairs takes place at much lower
cross-sections than single W or Z production. While we expect of the
order of 100,000 reconstructed Z bosons per experiment per leptonic
channel in one femtobarn of data, the expected yield for diboson
processes extends down to about one event per femtobarn for ZZ
production. Main emphasis of diboson reconstruction at this stage is
therefore establishing signals and measuring the absolute absolute
cross-section.

Very interesting results can be obtained from a measurement of the
Z$\gamma$ production rate. Since there are no ZZ$\gamma$ or
Z$\gamma\gamma$ vertices in the Standard Model, Z$\gamma$ combinations
can only be produced by initial state or final state radiation. Any
additional contributions would indicate new physics.
CDF\cite{cite-cdfzgammawgamma} and D\O\cite{cite-d0zgamma} investigate
Z$\gamma$ production in Z$\to$ee final states with a photon of at
least 7$\,$GeV. Photons from initial state and final state radiation
can be distinguished by looking at the three-body ee$\gamma$ mass in
addition to the ee mass.  Both experiments find agreement of the
observed production rate with Standard Model predictions, and in
particular no deviation from the expectation is observed at large
photon transverse energies or in the ISR/FSR distributions.

W$\gamma$ production does have a leading order contribution. Both
experiments measure cross-sections in good agreement with the standard
model prediction. CDF\cite{cite-cdfzgammawgamma} does this measurement
in the W$\to\mu\nu$ channel, whereas D\O\cite{cite-d0wgamma} uses both
electron and muon final states and employs a very stringent final
state radiation veto by requiring the W$\gamma$ three-body mass to
exceed 110$\,$GeV. D\O~increases sensitivity to anomalous couplings by
studying the charge signed rapidity difference $Q_{\ell}\times
\left[y(\gamma)-y(\ell)\right]$, which is expected to vanish at zero
for the Standard Model. This measurement will clearly unfold its full
potential once larger data samples are available.

\begin{figure}
\begin{center}
\epsfig{file=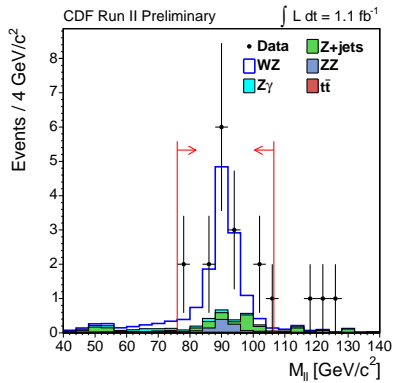,height=5cm}
\caption{\label{fig-cdfwz}\it Dilepton mass distribution of CDF WZ candidate events.}
\end{center}
\end{figure}

The production rates of massive boson pairs WW, WZ and ZZ are
predicted to spread over an order of magnitude in a similar range as
other important processes such as top quark pair production. While the
signal of WW production has been clearly established
(see\raisebox{-0.8ex}{\cite{cite-cdfww}} for a recent CDF
measurement), the WZ state is just barely accessible to observation
now. All combinations of electrons and muons in the final state are
considered to maximise reconstruction efficiency. CDF\cite{cite-cdfwz}
finds 16 WZ candidates in their approximately 1$\,$fb$^{-1}$ data
sample (see Fig.~\ref{fig-cdfwz}), with an expected background
contribution of $2.65\pm0.28\pm0.33\pm0.09$ events. This six standard
deviation excess above the background expectation constitutes the
first observation of the WZ channel.  D\O\cite{cite-d0wz} did a
similar analysis, but due to a combination of various small effects
their signal of 12 events including $3.61\pm0.20$ expected background
events remains below the formal threshold for an observation. Both
experiments do measure cross-sections in good agreement with the
Standard Model.

The lowest end of the diboson production cross-section spectrum, ZZ
production with an expected Standard Model cross-section of
$1.4\pm0.1\,$pb is hardly accessible to the Tevatron experiments so
far. CDF\cite{cite-cdfzz} performed a search for this channel,
finding one candidate event where approximately two are expected on
average. They can therefore quote a cross-section upper limit of
3.8$\,$pb at 95\% C.L. It is reasonable to expect that the ZZ channel
will be observed at the Tevatron once its full Run II dataset is
becoming available.

An overview over recent diboson results is given in Table~\ref{tab-xsec}.

\section{Discussion}
Leptonic final states of W and Z bosons are fairly clear signatures
even within the large background associated with hadron colliders.
High cross-section processes like single W or Z production therefore
provide an ideal laboratory for precision studies of parton density
functions. Rare electroweak processes like production of massive boson
pairs can already be identified down to cross-sections smaller than
top quark pair production. While we can realistically expect to
observe signatures like Z pairs (predicted at 1.4$\pm$1$\,$pb) at the
Tevatron with a 4--8$\,$fb$^{-1}$ data sample per experiment, it seems
unlikely that signals much smaller than that can be identified
directly, such as a hypothetical Standard Model H$\to$WW contribution
at an expected cross-section another order of magnitude below that of
ZZ production.

\section{Acknowledgements}
I thank the personnel involved in operating the Tevatron and the
D\O~and CDF detectors, as well as the people dedicating their time to
providing the software infrastructure for data handling, event
reconstruction and physics analysis. I would also like to express my
gratitude to the CDF and D\O~electroweak physics groups and their
conveners for providing me with input for this conference.

\end{document}